\documentclass[a4paper, 12pt]{article}
\usepackage[russian]{babel}
\usepackage[cp866]{inputenc}
\usepackage{graphicx}
\usepackage{amsfonts}
\usepackage{amsmath}
\usepackage{amssymb}

\def\np{{\vfill\eject}}

\begin{document}

\baselineskip 8.0mm
\thispagestyle{empty}

\centerline{\large\bf Relativistic effects and dark matter in the
solar system} \centerline{\large\bf from observations of planets and
spacecrafts}
\bigskip
\centerline{\bf E.V. Pitjeva$^{1*}$, N.P. Pitjev$^{2**}$}
\centerline{\it $^1$Institute of Applied Astronomy of RAS, St. Petersburg, Russia}
\centerline{\it $^2$St. Petersburg State University, St. Petersburg, Russia}
\centerline{\small Accepted 2013}

\bigskip\noindent

The high precision of the latest version of the planetary ephemeris
EPM2011 enables one to explore more accurately a variety of small
effects in the solar system. The processing of about 678 thousand of
position observations of planets and spacecrafts for 1913--2011 with
the predominance of modern radar measurements resulted in improving
the PPN parameters, dynamic oblateness of the Sun, secular variation
of the heliocentric gravitational constant $GM_{\odot}$, and
variation range of the gravitational constant $G$.
This processing made it possible to estimate the potential
additional gravitational influence of dark matter on the motion of
the solar system bodies. The density of dark matter ${\rho}_{dm}$,
if any, turned out to be substantially below the accuracy achieved
by the present determination of such parameters. At the distance of
the orbit of Saturn the density ${\rho}_{dm}$ is estimated to be
under $1.1\cdot10^{-20}$ g/cm$^3$, and the mass of dark matter in
the area inside the orbit of Saturn is less than $7.9\cdot10^{-11}$
M$_{\odot}$ even taking into account its possible tendency to
concentrate in the center.


\vskip 3em \noindent

\vfill\noindent

$^*$E-mail:    {\tt evp@ipa.nw.ru}

$^{**}$E-mail: {\tt ai@astro.spbu.ru}

\newpage
\centerline{\bf 1. INTRODUCTION}
\smallskip


  The possibility to test and refine various relativistic and cosmological
effects from the analysis of the motion of the solar system bodies
is due to the present meter accuracy radio techniques (Standish,
2008) and millimeter accuracy laser techniques (Murphy et al., 2008)
for the distance measurements. Just these techniques have provided
an observational foundation of the contemporary high--precision
theories of planetary motions.

  The numerical theories of planetary motions have been improved and developed by
several groups in different countries and their accuracy is
constantly growing. The progress is related with the increase of the
number of high--precision radio observations and the inclusion of a
number of small effects (perturbations from a set of asteroids, the
solar oblateness perturbations, etc.) in constructing a dynamic
model of the solar system. The radio technical observations, having
much higher accuracy as compared with the optical ones, are commonly
used now in astrometric practice. These high--precision measurements
covering more than 50 year time interval allow us to find the
orbital elements, masses and other parameters determining the motion
of the bodies. Moreover, they also give a possibility to check some
relativistic parameters, to estimate the secular change of the
heliocentric gravitation constant and to examine the presence of
dark matter in the solar system. The last point is of particular
importance for the contemporary cosmological theories. A more
accurate and extensive set of observations permits us not only to
determine the relativistic perihelion precession of planets, but
also to estimate the oblateness of the Sun with the corresponding
contribution into the drift of the perihelia. Moreover, these
observations provide a means for finding the secular variation of
the heliocentric gravitational constant $GM_{\odot}$  and the
constraint on the secular variation of the gravitational constant
$G$  (Pitjeva \& Pitjev, 2012). In addition, these precise
observations enable us to consider the assumption of the presence of
dark matter in the solar system and to estimate the upper limits of
its mass and density.


 The present research has been performed on the basis of the current version of  EPM2011
(the numerical Ephemerides of the Planets and the Moon) of IAA RAS.

\bigskip
\centerline{\bf 2. THE PLANETARY EPHEMERIS EPM2011}
\smallskip

  Numerical Ephemerides of Planets and the Moon (EPM) had started
in the 1970s. Each subsequent version is characterized by additional
new observations, refined values of the orbital elements and masses
of the bodies, an improved dynamical model of the celestial bodies
motion, as well as a more advanced reduction of observational data.


All presently used main planetary ephemerides DE (Standish, 1998),
EPM (Pitjeva, 2005a), and INPOP (Fienga,  2008) are based on General
Relativity involving the relativistic equations of celestial bodies
motion and light propagation as well as the relativistic time
scales. In addition, these ephemerides involve estimating from
observations some parameters ($\beta$, $\gamma$, $\dot G$) to check
their compatibility with General Relativity.


The current EPM2011 ephemerides were constructed using approximately
680 thousand data (1913-2011) of different types. The equations of
the bodies motion were taken within the parameterized
post--Newtonian n--body metric in the barycentric coordinate system
-- BCRS (Brumberg 1991), the same as that of DE. Integration in TDB
(Barycentric Dynamical Time) time scale (see the IAU2006 resolution
B3) was performed using the Everhart's method over the  400-year
interval (1800-2200) with the lunar and planetary integrator of the
ERA-7 software package (Krasinsky $\&$ Vasilyev,  1997). The EPM
ephemerides including also the time differences TT-TDB, and seven
additional asteroids, namely, Ceres, Pallas, Vesta, Eris, Haumea,
Makemake, Sedna, are available via FTP by means of
ftp://quasar.ipa.nw.ru/incoming/EPM/.


  Since the basic observational data for producing the next version of the
planetary ephemerides EPM2011 were mainly related to the
spacecrafts, the control of the orientation of the  EPM2011
ephemerides with respect to the ICRF frame has required a particular
attention. For this purpose we have used the VLBI observations of
the spacecrafts near planets at the background of quasars. The
coordinates of the quasars are given in the ICRF frame (Table 1),
where ($\alpha+\delta$) are one--dimensional measurements of the
$\alpha$ and $\delta$ combination, ($\alpha,\delta$) being the
two--dimensional measurements (the position of the planet is
observed to be displaced relative to the base ephemeris (DE405) by a
correction measured counter-clockwise along a line at an angle to
 the right ascension axis, see Folkner, 1992).

\np

{\bf Table 1.} VLBI observations of near--planet spacecrafts at the
ICRF background quasars
\smallskip
\begin{tabular}{|c|c|c|c|}
\noalign{\smallskip}
\hline
\noalign{\smallskip}
 Planet & Spacecraft & Interval of observations & Number of observations \\
\noalign{\smallskip}
\hline
\noalign{\smallskip}
Venus & Magellan & 1990-1994 & 18($\alpha+\delta$) \\[-2pt]
\noalign{\smallskip}
       & Venus Express & 2007-2010 & 29($\alpha+\delta$) \\[-2pt]
\noalign{\smallskip}
\hline
\noalign{\smallskip}
Mars & Phobos & 1989 & 2($\alpha+\delta$) \\[-2pt]
\noalign{\smallskip}
     & MGS & 2001-2003 & 15($\alpha+\delta$) \\[-2pt]
\noalign{\smallskip}
     & Odyssey & 2002-2010 & 86($\alpha+\delta$) \\[-2pt]
\noalign{\smallskip}
     & MRO & 2006-2010 & 41($\alpha+\delta$) \\[-2pt]
\noalign{\smallskip}
\hline
\noalign{\smallskip}
Saturn & Cassini & 2004-2009 & 22($\alpha,\delta$) \\[-2pt]
\noalign{\smallskip}
\hline
\noalign{\smallskip}
\end{tabular}

\smallskip

The accuracy of such observations increased to tenths of mas (1 mas
= 0".001) for Mars and Saturn in 2001-2010 (Jones et al.,
2011)enabling us to improve the orientation of EPM ephemerides
(Table 2) in the same way, as it was done by Standish (1998). The
angles of rotation of the Earth-Moon barycenter vector about the
x,y,z-axes of the BCRS system were obtained from VLBI observations
described above.


{\bf Table 2.} The angles of rotation of the EPM2011 ephemerides to ICRF
(1 mas = $0\rlap.''001$)
\smallskip
\begin{tabular}{|c|c|c|c|c|}
\noalign{\smallskip}
\hline
\noalign{\smallskip}
Interval of & Number of & $\varepsilon_x$ & $\varepsilon_y$ & $\varepsilon_z$ \\
observations & observations & mas & mas & mas \\
\noalign{\smallskip}
\hline
\noalign{\smallskip}
1989-2010 & 213 & $-0.000 \pm 0.042$ & $-0.025 \pm 0.048$ & $0.004 \pm 0.028$ \\[-2pt]
\hline
\noalign{\smallskip}
\end{tabular}

\smallskip

More than 270 parameters are estimated in the planetary part of
EPM2011 ephemerides as follows:

- the orbital elements of planets and satellites of the outer planets,

- the value of the astronomical unit or $GM_{\odot}$,

- the angles of orientation of the EPM ephemerides with respect to the ICRF system,

- parameters of the Mars rotation and the coordinates of the three
Mars landers,

- masses of 21 asteroids, the average density of the taxonomic class of
asteroids (C, S, M),

- the mass and radius of the asteroid ring and the mass of the TNO ring,

- the mass ratio of the Earth and Moon,

- the quadrupole moment of the Sun and the solar corona parameters
for different conjunctions of the planets with the Sun,

- the coefficients for the Mercury topography and the corrections to the level
surfaces of Venus and Mars,

- coefficients for the additional phase effect of the outer planets.

In the lunar part of EPM ephemerides about 70 parameters are
estimated from LLR data (see for example, Krasinsky, Prokhorenko,
Yagudina, 2011). All estimated parameters in both parts are
consistent within the frame of the combined theory of motion of the
planets and the Moon given by the EPM ephemerides.













  The initial parameters of EPM2011 represent the constants adopted by the IAU GA 27
(Luzum et al., 2011) as the current best values for ephemeris
astronomy. Among them five constants are resulted from the ephemeris
improvement of DE and EPM ephemerides (Pitjeva \& Standish, 2009).
These five parameters adjusted from processing all observations for
EPM2011 are as follows:

the masses of the largest asteroids, i.e. $M_{Ceres} / M_{\odot}$ =
$4.722(8)\cdot10^{-10}$, $M_{Pallas} / M_{\odot}$  =
$1.047(9)\cdot10^{-10}$, $M_{Vesta} / M_{\odot}$  =
$1.297(5)\cdot10^{-10}$;

ratio of the masses of the Earth and Moon
 $M_{Earth}/M_{Moon} = 81.30056763 \pm 0.00000005 $;

the  value of the astronomical unit in meters $ au =
(149597870695.88 \pm 0.14)$  or the heliocentric gravitation
constant $GM_{\odot} = (132712440031 \pm 1)$ km$^3$/s$^2$.

  Presently, in accordance with the IAU 2012 resolution B2 the astronomical unit
(au) is re-defined by fixing its value. Up to now, both values of au
and the heliocentric gravitation constant ($GM_{\odot}$) were in
use.  It was possible to determine the au value and to calculate the
value of $GM_{\odot}$ from it, or vice versa, to determine
$GM_{\odot}$ and to calculate the value of au from it. Here the
values of au and $GM_{\odot}$ are given as in the paper Pitjeva \&
Standish, (2009) published before the IAU 2012 resolution B2. At
present, only the value of $GM_{\odot}$ is estimated from
observations.


A serious problem in developing modern planetary ephemerides arises
due to the necessity to take into account the perturbations caused
by asteroids.  The factors affecting the planetary motions and
needed to be included in developing high--precision ephemerides are
of particular consideration in this paper. To start with, the hazard
near--Earth asteroids are relatively small (D < 5 km) and their
perturbations do not affect practically the Earth motion. That's why
they are not examined in this paper. The main asteroid belt
substantially affecting the motion of Mars and some other planets is
modeled in the EPM ephemerides by using the motion of 301 large
asteroids and a homogeneous material ring representing the influence
of all other numerous small asteroids (Krasinsky et al., 2002;
Pitjeva, 2010a). The parameters characterizing the ring of small
asteroids (its mass and radius) were determined from the analysis of
observations resulting in the values:
$$ M_{ring}=(1.06\pm1.12)\cdot10^{-10} M_{\odot} \,\, (3\sigma), \quad
R_{ring}=(3.57\pm0.26)\,\, (3\sigma) \,\, \hbox{au}. $$


 The total mass of the asteroid main belt represented by the sum of the mass of
301 largest asteroids and the homogeneous material ring (involving
the main uncertainty) is
\begin{equation}\label{f-1}
 M_{belt} = (12.29 \pm 1.13)\cdot10^{-10} M_{\odot}\,\, (3\sigma), \hbox{that is}
\approx 3 M_{Ceres}.
\end{equation}
This value of  $M_{belt}$ is close to $M_{belt} = (13.3 \pm
0.2)\cdot10^{-10} M_{\odot} \, (\sigma), $ obtained from the Mars
ranging data in the paper by Kuchynka \& Folkner (2013) by means of
another method in estimating the masses of 3714 individual
asteroids.


Hundreds of trans--neptunian objects (TNO) discovered in recent
years also affect the motion of the planets, especially outer ones.
A dynamic model of EPM ephemerides includes Eris (the planet--dwarf
found in 2003 and surpassing Pluto by its mass) and other 20 largest
TNO into the process of the simultaneous integration. Perturbations
from  other TNO are modeled by the perturbation from a homogeneous
ring located in the ecliptic plane with the radius of 43 au and the
 mass estimated in (Pitjeva, 2010a). The mass of the TNO ring found from
the analysis of observations amounts to
$$ M_{TNOring} = (501 \pm 249)\cdot10^{-10} M_{\odot}\,\, (3\sigma).$$
 The total mass of all TNOs including the mass of Pluto, 21 largest TNOs
and the TNO ring comes to
\begin{equation}\label{f-2}
M_{TNO} = (790 \pm 250)\cdot10^{-10} M_{\odot} \qquad (3\sigma),
\hbox{ that is}  \approx 164 M_{Ceres}  \hbox{ or }  \approx 2 M_{Moon}.
\end{equation}
  In addition to the mutual perturbations of the major planets and the Moon the
EPM2011 dynamic model includes

- the perturbations of the 301 most massive asteroids,

- the perturbations from the remaining minor planets of the main asteroid belt
modeled by a homogeneous ring,

- the perturbations from the 21 largest TNO,

- the perturbations from the remaining TNOs modeled by a uniform
ring at the average distance of 43 au,

- the relativistic perturbations,

- the perturbation due to the oblateness of the Sun estimated in
EPM2011 fitting as ($J_2=2\cdot10^{-7}$).








\bigskip
\centerline{\bf 3. OBSERVATION DATA AND THEIR REDUCTIONS}
\smallskip

 The total amount of the high-precision observations used for fitting EPM2011 has been
increased due to the recent data. They include 677 670 positional
measurements of different types for 1913-2011 from classic meridian
measurements to modern spacecraft tracking data (Table~3).


{\bf Table 3.} The observational material
\smallskip

\begin{tabular}{|l|c|c|c|c|}
\noalign{\smallskip} \hline \noalign{\smallskip}
Planet &\multicolumn{2}{|c|}{Radio observations} &\multicolumn{2}{|c|}{Optical observations}\\
\cline{2-5}
 & Time interval & Number & Time interval & Number \\
\noalign{\smallskip} \hline \noalign{\smallskip}
Mercury & 1964-2009 & 948 & -- & -- \\[-2pt]
\noalign{\smallskip} \hline \noalign{\smallskip} Venus  & 1961-2010
& 40061 & -- & -- \\ [-2pt] \noalign{\smallskip} \hline
\noalign{\smallskip}
Mars   & 1965-2010 & 578918 & -- & -- \\[-2pt]
\noalign{\smallskip} \hline \noalign{\smallskip}
Jupiter+4 sat. & 1973-1997 & 51 & 1914-2011 & 13364 \\[-2pt]
\noalign{\smallskip} \hline \noalign{\smallskip}
Saturn+9 sat. & 1979-2009 & 126 & 1913-2011 & 15956 \\[-2pt]
\noalign{\smallskip} \hline \noalign{\smallskip}
Uran+4 sat. & 1986 & 3 & 1914-2011 & 11846 \\[-2pt]
\noalign{\smallskip} \hline \noalign{\smallskip}
Neptun+1 sat. & 1989 & 3 & 1913-2011 & 11634 \\[-2pt]
\noalign{\smallskip} \hline \noalign{\smallskip}
Pluton & -- & -- & 1914-2011 & 5660 \\[-2pt]
\noalign{\smallskip} \hline
\noalign{\smallskip}
Total &  & 620110 & & 57560 \\
\noalign{\smallskip} \hline \noalign{\smallskip}
\end{tabular}

\smallskip

  Radar measurements (the detailed description of them is given in
Pitjeva 2005a, 2013) have a high accuracy. At present, the relative
accuracy $\sim 10^{-12}$ for the spacecraft trajectory measurements
became usual, exceeding the accuracy of classical optical
measurements by five orders of magnitude. However, in general only
Mercury, Venus, and Mars are provided with radio observations.
Initially, the surfaces of these planets were radio located from
1961 to 1995. Later on  many spacecrafts passed by, orbited or
landed to these planets. A large portion of the spacecraft data was
used to get the astrometric positions. There are much less radio
observations for Jupiter and Saturn, and only one set of the
three--dimensional normal points ($\alpha$, $\delta$, $R$) obtained
from the Voyager-2 spacecraft are available for Uranus and Neptune.
Therefore, the optical observations are still of great importance
for the outer planets. Thereby, the varied data of 19 spacecrafts
were used for constructing the EPM2011 ephemerides and estimating
the relevant parameters, in particular, the additional perihelion
precessions of the planets (see Table 4 below).


 The recent data from the spacecrafts have been added
to the previous ones for the latest version of the EPM ephemerides.
It involves data related to Odyssey, MRO (Mars Reconnaissance
Orbiter) (Konopliv et al., 2011), Mars Express (MEX), Venus Express
(VEX), and, more specifically, VLBI observations of Odyssey and MRO,
three--dimensional normal points of Cassini and Messenger
observations, along with the CCD observations of the outer planets
and their satellites obtained at Flagstaff and Table Mountain
observatories. The most part of observations was taken from the
database of JPL/Caltech created by Dr. Standish and continued by Dr.
Folkner.  MEX and VEX data provided by ESA became available thanks
to a kindness of Dr. Fienga   (private communications of T.Morlay to
A.Fienga).

\smallskip

The detailed description of methods for all reductions of planetary
observations (both optical and radar ones) was given  by Standish
(1990). This is a basic paper in the field of planetary observations
discussion. In the EPM ephemerides some reductions changed slightly
are described in Pitjeva 2005a, 2013. All necessary reductions
listed therein were introduced into actual observation data as
follows:
\bigskip

Reductions of the radar observations
\begin{itemize}
\itemsep -3mm
\item the reduction of moments of observations to a uniform time scale;
\item the relativistic corrections -- the time--delay of propagation of radio signals
in the gravitational field of the Sun, Jupiter and Saturn (Shapiro
effect) and the reduction of TDB time (ephemeris argument) to the
observer's proper time;
\item the delay of radio signals in the troposphere of the Earth;
\item the delay of radio signals in the plasma of the solar corona;
\item the correction for the topography of the planetary surfaces (Mercury, Venus,
Mars).
\end{itemize}
\bigskip
\np
  Reductions of the optical observations
\begin{itemize}
\itemsep -3mm
\item the transformation of observations to the ICRF frame

 catalogue differences => FK4 => FK5 => ICRF;
\item the relativistic correction for the light bending of the Sun;
\item the correction for the additional phase effect.
\end{itemize}




\bigskip
\centerline{\bf 4. RESULTS}
\smallskip

\bigskip
\centerline{\bf 4.1 Estimates of relativistic effects}
\smallskip

Some small parameters are determined  (in addition to the orbital
elements of the planets) while constructing  the EPM ephemerides
using new observations and the method
 similar to (Pitjeva, 2005a,b).  In
most cases the parameters can be found from the analysis of the
secular changes of the orbital elements. Therefore, the
uncertainties of their determination decrease with increasing the
time interval of observations.


The simplified relativistic equations of the planetary motion were
derived more than 30 years ago in different coordinate systems of
the Schwarzschild metric supplemented with coordinate parameter
alpha to specify  standard, harmonic, isotropic, or any other
coordinates. These equations were described in (Brumberg, 1972,
1991). For example, the integration exposed in  (Oesterwinter \&
Cohen, 1972) was made in the standard coordinates (alpha=1).
However, planetary coordinates turned out to be essentially
different for the standard and harmonic systems. It was shown in
(Brumberg, 1979)  that the ephemeris construction and processing
 of observations should be done in the same coordinate system resulting to
 the relativistic effects not dependent on the coordinate system (effacing of parameter alpha).
Later on, the resolutions of IAU (1991, 2000)
 recommended to use the harmonic coordinates for BCRS. In accordance with
the IAU 2000 resolution B1.3 modern planetary ephemerides should be
constructed in the harmonic coordinates for BCRS -- the barycentric
(for the solar system) coordinate system.


The parameters of the PPN formalism $\beta, \ \gamma$ used to
describe the metric theories of gravity must be equal to 1 in
General Relativity. The values of parameters $\beta, \gamma$  were
obtained simultaneously by using the EPM2011 ephemerides and the
updated database of high--precision observations (Table 3) to get
the relativistic  periodic and secular variations of the orbital
elements, as well as the Shapiro effect. Certainly, the periodic
variations of the orbital elements are smaller than the secular ones
but they are of importance to compute the planetary motion. We
derived expressions for the partial derivatives of the orbital
elements with respect to $\beta$ and $\gamma$
 using the analytical formulas for the relativistic perturbations
of the elements, including the secular and principal periodic terms
given in (Brumberg, 1972). This technique enabled us to get actually
 the values for $\beta$ and $\gamma$. Moreover, in the eighties of the last
 century we tested relativistic effects by processing the observations
available at that time. It turned out that the relativistic
ephemeris for any observed planet provided a considerably better fit
of
 observations (by 10\%) than the Newtonian theory even if the latter
incorporated the observed perihelion secular motion (Krasinsky et
al., 1986).


The obtained values of  $\beta$ and $\gamma$ read
\begin{equation}\label{f-3}
 \beta -1 =-0.00002 \pm 0.00003,\quad \gamma -1 = +0.00004 \pm 0.00006 \quad
(3\sigma).
\end{equation}
The uncertainties in (3) significantly decreased as compared with
the results for the EPM2004 (Pitjeva, 2005b) and EPM2008 (Pitjeva,
2010b)
$$ |\beta - 1|<0.0002, \ |\gamma - 1|<0.0002. $$

In  (Fienga et al., 2011) based on the INPOP10a planetary ephemeris
 these parameters were determined separately fixing one of these
two values, either  $\beta = 1$, or $\gamma = 1$. Yet the ephemeris
fitting results in
$$\beta -1 = -0.000062 \pm 0.000081, \quad \gamma -1 = +0.000045 \pm 0.000075. $$

For comparison we also quote the new $\gamma$ value obtained by
using Very Long Baseline Array measurements of radio sources by
Fomalont et al., 2009, i.e. $\gamma = 0.9998 \pm 0.0003$.

All the obtained values of $\beta, \gamma$ are in the close vicinity
of 1 within the limits of their uncertainties. As the uncertainties
of these parameters decrease, the range of possible values of the
PPN parameters narrows, imposing increasingly stringent constraints
on the gravitation theories alternative to General Relativity.





\bigskip
\centerline{\bf 4.2 Estimations of the solar dynamic oblateness}
\smallskip

  The solar oblateness produces the secular trends in all elements of the
planets with the exception of their semimajor axes and
eccentricities (see, for example, Brumberg, 1972). Therefore, the
dynamic solar oblateness can be determined together with other
parameters from observations in constructing a theory of planetary
motion. The quadrupole moment of the Sun characterizing the solar
oblateness was found in EPM2011 to be
\begin{equation}\label{f-4}
   J_2 = (2.0 \pm 0.2)\cdot10^{-7} \quad (3\sigma),
\end{equation}
that is close to the previous result (Pitjeva, 2005a,b) for EPM2004
   $ J_2 = (1.9 \pm 0.3)\cdot10^{-7} $ and the result of INPOP10a
(Fienga et al., 2011) $J_2 =(2.40  \pm 0.25)\cdot10^{-7} \quad
(1\sigma).$


\bigskip
\centerline{\bf 4.3 Estimations of the secular changes of $GM_{\odot}$ and $G$}
\smallskip

The value of the secular change of the heliocentric gravitational
constant $GM_{\odot}$ has been updated for the expanded database and
the improved dynamical model of planetary motions (EPM2011). The
determination of secular variation $GM_{\odot}$ was carried out by
the method exposed in detail in (Pitjeva \& Pitjev, 2012) dealing
with the EPM2010 planetary ephemerides.


The  $GM_{\odot}$ change was determined by the weighted method of
the least squares with all the basic parameters of the EPM2011
ephemerides. In determining $\dot{GM}_{\odot}$, it was taken into
account that the acceleration between the Sun and any planet varies
with time when $GM_{\odot}$ is changing, but the acceleration
between any two planets remains unchanged. This is different from
the situation when one looks for the $G$ change involving the
corresponding change of the accelerations of all bodies.  It should
be noted that when we determine $\dot{G}$ using the planetary
motions (Pitjeva \& Pitjev, 2012), it is the Sun that contributes
most of all. Indeed, the equations of planetary motion include the
products of the masses of bodies and the gravitational constant, the
main term exceeding other terms by several orders of magnitude is
that for the Sun ($GM_{\odot}$). Therefore, as the $GM_{\odot}$ term
dominates, it is impossible to separate the change of $G$ from the
change of $GM_{\odot}$  considering only the motion of the planets
(Pitjeva \& Pitjev, 2012).
   However, if the change of the solar mass ($M_{\odot}$) may be estimated from
the independent astrophysical data, then based on the change of the
$GM_{\odot}$ and the limits of the $M_{\odot}$ change, the limits of
the gravitation constant (G) change can be obtained taking into
account the following relation
\begin{equation}\label{f-5}
\dot {GM_{\odot}}/GM_{\odot}=\dot G/G+\dot M_{\odot}/M_{\odot}.
\end{equation}
(see details in Pitjeva \& Pitjev, 2012).


  Contrary to $\dot G$, it is the change of $GM_{\odot}$ that can be determined more accurately and
reliably using the planetary motions. To control the stability of
the solution for $\dot{GM}_{\odot}$ and to obtain the more reliable
error, we considered various fitting versions with different numbers
of the parameters (the number of the adjusted masses of asteroids,
perihelion precessions, etc.). The time--decrease of $GM_{\odot}$
was found to be
\begin{equation}\label{f-6}
 \dot{GM}_{\odot}/GM_{\odot} = (-6.3 \pm 4.3)\cdot10^{-14} \quad \hbox{per year}
\quad (2\sigma).
\end{equation}
This decrease is caused by the loss of the solar mass $M_{\odot}$
through radiation and the solar wind. The estimate of the
uncertainty for this value is  more reliable and larger than in
(Pitjeva \& Pitjev, 2012). Analysis of versions with different
numbers of the fitting parameters demonstrates that the value of
$\dot {GM}_{\odot}$ and its uncertainty are the most sensitive to
the parameters related to the main asteroid belt, i.e. the amount of
the adjusted masses of the selected large asteroids and the
estimated characteristics of the ring representing the effect of the
small asteroids. The value obtained by Folkner (Konopliv et al.,
2011) for DE423 ephemerides from the Mars ranging data is
$$ \dot{GM}_{\odot}/GM_{\odot} = (1 \pm 16)\cdot10^{-14} \quad
\hbox{per year}. $$ The uncertainty of this value is larger (and may
be more reliable) due to taking into account the uncertainties of
many other asteroid masses remained unestimated.


Estimation of $\dot{M}_{\odot}$ has been made  by means of the
astrophysical data using the values for the average solar radiation
and solar wind, and amount of comet and asteroid matter falling on
the Sun. The obtained limits of the possible change of $M_{\odot}$
can be bounded by the inequality (Pitjeva \& Pitjev, 2012)
\begin{equation}\label{f-7}
 -9.8\cdot10^{-14} < \dot{M}_{\odot}/M_{\odot} < -3.6\cdot10^{-14} \quad
\hbox{per year}.
\end{equation}
This interval may be narrowed due to the more accurate estimation of
matter falling on the Sun. From (6) and taking into account (5) and
(7), the $\dot{G}/G$ value is found to be within the interval (with
the 95$\%$ probability)
\begin{equation}\label{f-8}
 -7.0\cdot10^{-14} < \dot{G}/G < +7.8\cdot10^{-14} \quad \hbox{per year}.
\end{equation}
The interval (8) imposes the more rigid limits on the possible
change of $G$ than the results of the determination $\dot{G}$
obtained from processing lunar laser observations by Turyshev \&
Williams (2007)
$$\dot{G}/G = (6 \pm 7)\cdot10^{-13} \quad \hbox{per year}$$
and Hofmann, Muller \& Biskupek (2010)
$$\dot{G}/G = (-7 \pm 38)\cdot10^{-14} \quad \hbox{per year}.$$


\bigskip
\centerline{\bf 4.4 Estimations of dark matter in the solar system}
\smallskip

 It is proposed in the modern cosmological theories that the bulk of the average
density of the universe falls on dark energy (about 73\%) and the
dark matter 23\%, whereas the baryon matter contains about 4\%
(Kowalski et al., 2008). The nature of dark matter is non--baryon
and its properties are hypothetical (Bertone, Hooper \& Silk, 2005;
Peter, 2012).


  Despite the possible absence or the very weak interaction of dark matter with
ordinary matter, it must possess the capacity of gravity, and its
presence in the solar system can be manifested through its
gravitational influence on the body motion. Attempts to detect the
possible influence of dark matter on the motion of objects in the
solar system have already been made (Nordtvedt, Mueller \& Soffel,
1995; Anderson et al., 1989; Anderson et al., 1995; Sereno \&
Jetzer, 2006; Khriplovich \& Pitjeva, 2006; Khriplovich, 2007;
Frere, Ling \& Vertongen, 2008).


 The additional gravitational influence may depend on the density of dark
matter, its distribution in space, etc. We assume, as it is usually
done (Anderson et al., 1989; Anderson et al., 1995; Gron \& Soleng,
1996; Khriplovich \& Pitjeva, 2006; Frere et al., 2008), that  dark
matter is distributed in the solar system spherically symmetric
relative to the Sun. Then we may suppose that any planet at distance
$r$ from the Sun can be undergone an additional acceleration from
invisible matter along with the accelerations from the Sun, planets,
asteroids, trans--neptunian objects
$$   {\bf\ddot r}_{dm} = - {GM(r)_{dm} \over {r^3}} \bf r  ,$$
where $M(r)_{dm}$ is the mass of the additional matter in a sphere
of radius $r$ around the Sun.

 Testing the presence of the additional gravitational environment can be carried
out either by finding the additional acceleration, as was made, for
example, in (Nordtvedt et al., 1995; Anderson et al., 1989), or the
additional perihelion drift (for example, Gron \& Soleng, 1996).

 The first method determines actually if there is any extra mass inside
the spherically symmetric volume, in addition to the masses of the
Sun, planets and asteroids already taken into account. Any detected
correction to the central attracting mass (or to the heliocentric
gravitational constant $GM_{\odot}$) from the observational data
separately for each planet would result in its increased value in
accordance with the additional mass within the sphere with the mean
radius of the planetary orbit.




 The second way is related with an unclosed trajectory of motion
in the presence of the additional gravitational medium and the drift
of the positions of the pericenters and apocenters from revolution
to revolution in contrast to the purely Keplerian case of the
two-body problem. Denoting the integrals of energy and area by $E,
J$, and the spherically symmetric potential by $U(r)$ the equations
of motion of a unit mass along the radius $r$ and along the
azimuthal coordinate $\theta$ read, respectively, (Landau \&
Lifshitz, 1969)
\begin{equation}\label{f-9}
   \dot r = ( 2[E+U(r)] - J^2/r^2 )^{1/2} ,
\end{equation}
\begin{equation}\label{f-10}
    {d \theta \over dr} = {{J/r^2} \over ( 2[E+U(r)] - J^2/r^2 )^{1/2}} .
\end{equation}
In the Keplerian two-body problem the oscillation periods along the
radius $r$ (from the perihelion to the apocenter and back) and along
azimuth $\theta$ around the center coincide, and the positions of
the pericenter and apocenter are not displaced from revolution to
revolution. The additional gravitating medium leads to a shorter
radial period and a negative drift of the position of the pericenter
and apocenter (in a direction opposite to the planetary motion). The
perihelion precession for the uniformly distributed matter
(${\rho}_{dm}$=const) depends on the orbital semimajor axis $a$ and
eccentricity
 $e$ of the planetary orbit (Khriplovich \& Pitjeva, 2006)
\begin{equation}\label{f-11}
\Delta {\theta}_0 = -4{\pi}^2{\rho}_{dm}a^3(1-e^2)^{1/2}/M_{\odot},
\end{equation}
where $\Delta {\theta}_0$ is the perihelion drift for one complete
radial oscillation.

  Estimations of the density and mass of dark matter are produced often under the
assumption that it changes very slowly or is constant within the
solar system, i.e. under the assumption of the uniform distribution
of dark matter. A number of papers (Lundberg \& Edsjo, 2004; Peter,
2009; Iorio, 2010) assume the concentration of dark matter to the
center and even its capture and dropping on the Sun. The latter
assumption should be made with caution. In the item 4.3 (as well as
in Pitjeva \& Pitjev, 2012), it was found that the heliocentric
gravitational constant $GM_{\odot}$ decreases, so there is a
stringent limitation on the amount of possible dark matter dropping
on the Sun. The constraint on the possible presence of dark matter
inside the Sun (no more than 2-5\% of the solar mass) was also
obtained in  (Kardashev, Tutukov \& Fedorova, 2005), where the
physical characteristics of the Sun have been carefully analyzed.

   Both approaches have been applied in the present work. The more sophisticated
consideration is given in  (Pitjev \& Pitjeva, 2013).



 The corrections to the additional perihelion precession and to the central
mass were obtained by fitting the EPM2011 ephemerides to about 780
thousand of observations of the planets and spacecrafts (Table 3).
The fitting was done by the weighted method of the least squares.
The various test solutions differing from one another by the sets of
the adjusted parameters were considered for obtaining the reliable
values of these parameters and their uncertainties ($\sigma_i$) in
the same manner as for getting the $\dot{GM}_{\odot}$ estimation.

 The resulting values are exceeded by
their uncertainties ($\sigma$) indicating that the dark matter
density
 ${\rho}_{dm}$, if any, is very small being lower than the accuracy of these
parameters achieved by the modern determination. The obtained
opposite signs for the values $\Delta \pi$ and $\Delta M_{0}$ for
the various planets also show the smallness of such effects.



 The relative uncertainties in the corrections to the central mass
from the observations separately for each planet were significantly
greater than that for the additional perihelion precessions
exceeding the corrections to the central mass themselves in several
times or even by several orders of magnitude. It should be
remembered that the integral estimation of the dark matter mass
falling into a spherically symmetric (relative to the Sun) volume
depends on the accuracy of knowledge of all body masses into this
volume. Basically, it is due to the inaccurate knowledge of the
masses of asteroids.

More accurate results were obtained for estimates of the perihelion
precessions (Table 4) allowing to estimate the local density of dark
matter at the mean orbital distance of a planet. Here, the
uncertainties of determination of the corrections are comparable
with the values themselves. Therefore, the estimates from Table 4
were actually used.

The investigation of the additional perihelion precession of the
planets was carried out taking into account all other known effects
affecting the perihelion drift. Indeed, if there is an additional
gravitating medium, then a negative drift of the perihelion and
aphelion occurs from revolution to revolution in accordance with the
formula (11). Since the growth of the perihelion drift is
accumulated, this criterion can be sensitive enough for verifying
the presence of additional matter.





{\bf Table 4.} Additional perihelion precessions from observations
of planets and 19 spacecrafts

\begin{tabular}{|c|c|c|}
\noalign{\smallskip}
\hline
\noalign{\smallskip}
Planets & $\dot \pi$& |${\sigma_{\dot \pi} / \dot \pi}$| \\
        &  mas/yr   & \\
\noalign{\smallskip}
\hline
\noalign{\smallskip}
Mercury & $-0.020 \pm 0.030$& 1.5 \\[-2pt]
\noalign{\smallskip}
\hline
\noalign{\smallskip}
Venus  & $0.026 \pm 0.016$& 0.62 \\ [-2pt]
\noalign{\smallskip}
\hline
\noalign{\smallskip}
Earth  & $0.0019 \pm 0.0019$& 1.0 \\ [-2pt]
\noalign{\smallskip}
\hline
\noalign{\smallskip}
Mars   & $-0.00020 \pm 0.00037$& 1.9 \\[-2pt]
\noalign{\smallskip}
\hline
\noalign{\smallskip}
Jupiter & $0.587 \pm 0.283$& 0.48  \\[-2pt]
\noalign{\smallskip}
\hline
\noalign{\smallskip}
Saturn & $-0.0032 \pm 0.0047$& 1.5  \\[-2pt]
\noalign{\smallskip}
\hline
\noalign{\smallskip}
\end{tabular}
\bigskip

All the uncertainties of Table 4 are comparable or larger than the
absolute values obtained for perihelion precessions. These
uncertainties $\sigma_{\dot \pi}$ may be treated as the upper limits
for the possible additional drifts of the secular motion of the
perihelia, and can give the upper limit for the density of the
distributed matter by using (11). The resulting estimates
${\rho}_{dm}$ are shown in Table 5.


{\bf Table 5.} Estimates of the density ${\rho}_{dm}$ obtained from
$\sigma_{\Delta\pi}$ for the perihelion precessions

\begin{tabular}{|c|c|c|}
\noalign{\smallskip}
\hline
\noalign{\smallskip}
Planets & $\sigma_{\dot \pi}$ & ${\rho}_{dm}$ \\
        &  $''$/yr & g/cm$^3$ \\
\noalign{\smallskip}
\hline
\noalign{\smallskip}
Mercury & $0.000030$ & $<9.3\cdot 10^{-18}$\\[-2pt]
\noalign{\smallskip}
\hline
\noalign{\smallskip}
Venus & $0.000016$ & $<1.9\cdot 10^{-18}$\\[-2pt]
\noalign{\smallskip}
\hline
\noalign{\smallskip}
Earth  & $0.0000019$ & $<1.4\cdot 10^{-19}$\\ [-2pt]
\noalign{\smallskip}
\hline
\noalign{\smallskip}
Mars   & $0.00000037$ & $<1.4\cdot 10^{-20}$\\[-2pt]
\noalign{\smallskip}
\hline
\noalign{\smallskip}
Jupiter & $0.000283$ & $<1.7\cdot 10^{-18}$\\[-2pt]
\noalign{\smallskip}
\hline
\noalign{\smallskip}
Saturn & $0.0000047$ & $<1.1\cdot 10^{-20}$\\[-2pt]
\noalign{\smallskip}
\hline
\end{tabular}
\bigskip

The data based on the estimates  for the Earth, Mars and Saturn
yield the most stringent constraints on the density ${\rho}_{dm}$.
The high--precision series of observations of Saturn appeared when
the Cassini spacecraft arrived to it in 2004. There is the large and
long set of observations of Mars associated with many spacecrafts on
its surface and around it. The Earth orbit improvement is based on
all observations starting with the observations made from the Earth.
Assuming the homogeneous distribution ${\rho}_{dm}$ in the solar
system the most stringent constraint ${\rho}_{dm} < 1.1\cdot
10^{-20}$ g/cm$^3$ is obtained from the data for Saturn. Then the
mass $M_{dm}$ within the spherical volume with the size of Saturn's
 orbit is
\begin{equation}\label{f-12}
   M_{dm} < 7.1\cdot10^{-11} M_{\odot}.
\end{equation}
This value is about 2 times smaller than the uncertainty of the obtained total mass
of the main asteroid belt (1).


 Another version can be considered when a continuous medium has
some concentration to the center of the solar system. Investigations
under the assumption of density concentration to the center have
already been carried out, for example, by  Frere et al. (2008). We
have taken the model for ${\rho}_{dm}$ with the exponential
dependence on the distance $r$
\begin{equation}\label{f-13}
    {\rho}_{dm} = {\rho}_{0} \cdot e^{-cr} ,
\end{equation}
where ${\rho}_{0}$ is the central density and $c$ is a positive
parameter characterizing an exponential decrease of the density to
the periphery. The value of $c = 0$ corresponds to a uniform
density. Function (13) is everywhere finite and has no singularities
at the center and on the periphery. The mass inside a sphere of
radius $r$ for distribution (13) is
\begin{equation}\label{f-14}
 M_{dm} = 4\pi{\rho}_{0} \cdot {2 - e^{-cr}(c^2 r^2 + 2c r + 2) \over c^3} .
\end{equation}
In spite of the presence of $c^3$ in the  denominator this
expression does not have singularities for  $c \to 0$. The formula
(14) transforms  therewith into the expression for the mass of a
homogeneous sphere.


 The values in Table 5 may be considered as the limits of the density
${\rho}_{dm}$ at various distances. In a relatively narrow interval
of the radial distances caused by the eccentricity of the planetary
orbit the density of dark matter can be considered to be
approximately constant. The potential existence of the dark matter
$M_{dm}$ distributed between the Sun and the orbit of a planet gives
very small contribution (the tenths or elevenths fraction of the
magnitude) to the total attractive central mass determined by the
solar mass. Therefore, one can use the formula (11) and obtain the
local restrictive estimations for ${\rho}_{dm}$ in the neighborhood
of the planet orbit (Table 5).


With the assumption of the concentration to the center the estimate
of the mass of dark matter within the orbit of Saturn was determined
from the evaluation of the masses within the two intervals, i.e.
from Saturn to Mars and from Mars to the Sun. For this purpose the
most reliable data of Table 5 for Saturn
(${\rho}_{dm}<1.1\cdot10^{-20}$ g/cm$^3$), Mars (${\rho}_{dm}<1.4
\cdot10^{-20}$ g/cm$^3$) and Earth (${\rho}_{dm}<1.4\cdot10^{-19}$
g/cm$^3$) were used. Based on the data for Saturn and Mars a very
flat trend of the density curve (13) between Mars and Saturn was
obtained with $\rho_0 = 1.47\cdot 10^{-20}$ g/cm$^3$ and $c=0.0299$
au$^{-1}$. From these parameters the mass in the space between the
orbits of Mars and Saturn is $M_{dm} < 7.33\cdot10^{-11} M_{\odot}$.
 The obtained trend of the density curve (13) in the interval
between Mars and the Sun gives a steep climb to the Sun according to
the data for Earth and Mars with the parameters $\rho_0 =
1.17\cdot10^{-17}$ g/cm$^3$ and $c =4.42$ au$^{-1}$. For these
parameters the mass (14) between the Sun and the orbit of Mars is
$M_{dm} < 0.55\cdot10^{-11} M_{\odot}$.


 Summing masses for both intervals the upper limit for the total mass of
dark matter was estimated as $M_{dm} < 7.88\cdot10^{-11} M_{\odot}$
between the Sun and the orbit of Saturn, taking into account its
possible tendency to concentrate in the center. This value is less
than the uncertainty $\pm 1.13\cdot10^{-10} M_{\odot} \,(3\sigma)$
of the total mass of the asteroid belt. The value $M_{dm}$ does not
change perceptibly compared to the hypothesis of a uniform density
(12), although the trend of the density curve in the second case
provides the significant (by three orders of magnitude) increase to
the center.


\bigskip
\centerline{\bf CONCLUSION}
\smallskip
The estimations of the gravitational PPN parameters, the solar
oblateness, the secular change of the heliocentric gravitation
constant $ {GM}_{\odot}$ and the gravitation constant $G$, as well
as the possible gravitational influence of dark matter on the motion
of the planets in the solar system have been made on the basis of
the EPM2011 planetary ephemerides of IAA RAS using about 678,000
positional observations of planets and spacecrafts, mostly radio and
laser ranging ones.


 The PPN parameters turned out to be $\beta-1=-0.00002\pm0.00003,\quad \gamma-1=+0.00004\pm
0.00006 \quad (\sigma)$. Our estimation for the change of the
heliocentric gravitational constant is $ \dot{GM}_{\odot}/GM_{\odot}
= (-6.3 \pm 4.3)\cdot10^{-14} \quad \hbox{per year} \quad
(2\sigma)$. It was found also that the limits for the time variation
of the gravitational constant $G$ are $-7.0\cdot10^{-14} < \dot{G}/G
< +7.8\cdot10^{-14} \quad (2\sigma)$ per year.

 The mass and the level of dark matter density in the solar system, if any,
was obtained to be substantially lower than the modern uncertainties of these
parameters. The density of dark matter was found to be lower than
${\rho}_{dm}<1.1\cdot10^{-20}$ g/cm$^3$ at the distance of
the Saturn orbit, and the mass of dark matter in the area inside the
orbit of Saturn is less than $7.9\cdot10^{-11}$ M$_{\odot}$, even taking
into account its possible tendency to concentrate in the center.



\bigskip
\centerline{\bf REFERENCES}
\smallskip
{
\parindent = 0.0truecm

\hangindent = 0.7truecm
\hangafter = 1
Anderson J.D., Lau E.L., Taylor A.H., Dicus D.A., Teplitz D.A., Teplitz V.L.,
1989, ApJ, 342, 539

\hangindent = 0.7truecm
\hangafter = 1
Anderson J.D., Lau E.L., Krisher T.P., Dicus D.A., Rosenbaum D.C., Teplitz V.L.,
1995, ApJ, 448, 885

\hangindent = 0.7truecm
\hangafter = 1
Bertone G., Hooper D., Silk J., 2005, Phys. Rep. 405, 5-6, 279

\hangindent = 0.7truecm
\hangafter = 1
Brumberg V.A., 1979,  Celest. Mech., 20, 329

\hangindent = 0.7truecm
\hangafter = 1
Brumberg V., 1991, Essential relativistic celestial mechanics.
Adam Hilger, Bristol

\hangindent = 0.7truecm
\hangafter = 1
Fienga A., Manche H., Laskar J.,  Gastineau M., 2008, A\&A, 477, 315

\hangindent = 0.7truecm
\hangafter = 1
Fienga A., Laskar J., Kuchynka P., Manche H., Desvignes G., Gastineau M.,
Cognard I., Theureau G., 2011, CeMeDA, 111, issue 3, 363

\hangindent = 0.7truecm
\hangafter = 1
Folkner W.M., 1992, JPL IO 335.1-92-25. From
$http://iau-comm4.jpl.nasa.gov/plan-eph-data/vlbiformat_mar08.html$

\hangindent = 0.7truecm
\hangafter = 1
Fomalont E., Kopeikin S., Lanyi G., Benson J., 2009, ApJ, 699, issue 2, 1395

\hangindent = 0.7truecm
\hangafter = 1
Frere J.-M., Ling F.-S., Vertongen G., 2008, PhRvD, 77, Issue 8, id. 083005

\hangindent = 0.7truecm
\hangafter = 1
Gron O., Soleng H.H., 1996, ApJ, 456, 445

\hangindent = 0.7truecm
\hangafter = 1
Hofmann F., Muller J., Biskupek L., 2010, A\&A, 552, L5

\hangindent = 0.7truecm
\hangafter = 1
Iorio L., 2010, J. Cosmol. Astropart. Phys., 05, 018

\hangindent = 0.7truecm
\hangafter = 1
Jones D. L., Fomalont E., Dhawan V., Romney J., Folkner  W. M., Lanyi G.,
Border J., Jacobson R. A., 2011, AJ, 141, issue 2, 29

\hangindent = 0.7truecm
\hangafter = 1
Kardashev N.S., Tutukov A.V., Fedorova A.V., 2005, Astronomy Reports, 49,
issue 2, 134

\hangindent = 0.7truecm
\hangafter = 1
Khriplovich I.B., Pitjeva, E. V., 2006, IJMPD, 15, issue 04, 615

\hangindent = 0.7truecm
\hangafter = 1
Khriplovich I. B., 2007, IJMPD, 16, Issue 09, 1475

\hangindent = 0.7truecm
\hangafter = 1
Konopliv A. S., Asmar S. W., Folkner W. M., Karatekin O., Nunes D. C.,
Smrekar S. E., Yoder Ch. F., Zuber M. T., 2011, Icarus, 211, issue 1, 401

\hangindent = 0.7truecm
\hangafter = 1
Kowalski M., Rubin D., Aldering G., et al., 2008, ApJ, 686, 749

\hangindent = 0.7truecm
\hangafter = 1
Krasinsky G. A., Aleshkina E.Yu., Pitjeva E.V., Sveshnikov M.L., 1986,
in J. Kovalevsky, V.A. Brumberg, eds, Proc. IAU Symp. 114, Relativity in
celestial mechanics and astrometrym,  D. Reidel Publ.Com., Dordrecht, 315

\hangindent = 0.7truecm
\hangafter = 1
Krasinsky  G.A., Vasilyev M.V., 1997, in I.M. Wytrzyszczak, J.H. Lieske,
R.A. Feldman, eds, Proc. IAU Coll. 165, Dynamics and Astrometry of Natural and
Artificial Celestial Bodies, Dordrecht: Kluwer acad. Publ.,  239

\hangindent = 0.7truecm
\hangafter = 1
Krasinsky G.A., Pitjeva E.V., Vasilyev M.V., Yagudina E.I., 2002, Icarus, 158, 98

\hangindent = 0.7truecm
\hangafter = 1
Krasinsky G.A., Prokhorenko S.O., Yagudina E.I., 2011, in N. Capitaine, eds, Proc.
of Journees 2010 “Systemes de reference spatio-temporels, New challenges for
reference systems and numerical standards in astronomyine, Paris, France, 61

\hangindent = 0.7truecm
\hangafter = 1
Kuchynka P., Folkner W. M., 2013, Icarus, 222, 243

\hangindent = 0.7truecm
\hangafter = 1
Landau L.D., Lifshitz E.M., 1969, Mechanics,  Translated from the Russian by
J.B. Sykes and J.S. Bell. English 2d ed., Oxford, New York, Pergamon Press

\hangindent = 0.7truecm
\hangafter = 1
Lundberg J., Edsjo J., 2004, PhRvD, 69, 123505

\hangindent = 0.7truecm
\hangafter = 1
Luzum B. et al., 2011, CeMeDA, 110, Issue 4, 293

\hangindent = 0.7truecm
\hangafter = 1
Murphy T., Michelsen E., Orin A., Battat J., Stubbs C., Adelberger E., Hoyle C.,
Swanson H. 2008, in Kleinert H., Jantzen R.T., Ruffini R. eds, Proc. MG11
Meeting on General Relativity. Publ. by World Scientific Publishing Co. Pte.
Ltd., p. 2579

\hangindent = 0.7truecm
\hangafter = 1
Nordtvedt K.L., Mueller J., Soffel M., 1995, A\&A, 293L, L73

\hangindent = 0.7truecm
\hangafter = 1
Oesterwinter C., Cohen Ch.J., 1972, Celest. Mech., 5, N 3, 317

\hangindent = 0.7truecm
\hangafter = 1
Peter A., 2009, PhRvD 79(10), 103531

\hangindent = 0.7truecm
\hangafter = 1
Peter A., 2012, arXiv: astro-ph/1201.3942

\hangindent = 0.7truecm
\hangafter = 1
Pitjev N.P., Pitjeva E.V., 2013, Astronomy Letters, 39, issue 3, 141

\hangindent = 0.7truecm
\hangafter = 1
Pitjeva E.V., 2005a, Solar System Research, 39, issue 3, 176

\hangindent = 0.7truecm
\hangafter = 1
Pitjeva E.V., 2005b, Astronomy Letters, 31, issue 5, 340

\hangindent = 0.7truecm
\hangafter = 1
Pitjeva E.V., 2010a, in A.M. Finkelstein, W.F. Huebner, \& V.A. Shor, eds,
Proc. of the Intern. Conf. "Asteroid-Comet Hazard 2009", St. Petersburg:
Nauka, 237

\hangindent = 0.7truecm
\hangafter = 1
Pitjeva E.V., 2010b, in S.A. Klioner, P.K. Seidelman \& M.H. Soffel, eds,
Proc.IAU Symposium 261, Relativity in Fundamental Astronomy, 170

\hangindent = 0.7truecm
\hangafter = 1
Pitjeva E.V., 2013, Solar System Research, 47, issue 4, in print

\hangindent = 0.7truecm
\hangafter = 1
Pitjeva E.V., Pitjev N.P., 2012, Solar System Research, 46, issue 1, 78

\hangindent = 0.7truecm
\hangafter = 1
Pitjeva E.V., Standish E.M., 2009, CeMeDA, 103, 365

\hangindent = 0.7truecm
\hangafter = 1
Sereno M., Jetzer Ph., 2006, MNRAS, 371, 626

\hangindent = 0.7truecm
\hangafter = 1
Standish E.M., 1990, A\&A, 233, 252

\hangindent = 0.7truecm
\hangafter = 1
Standish E.M., 1998, Interoffice Memorandum, 312.F-98-048, 18p.

\hangindent = 0.7truecm
\hangafter = 1
Standish E.M., 2008, in AIP Conference Proceedings, 3rd Mexican Meeting on
Mathematical and Experimental Physics. Vol. 977, 254

\hangindent = 0.7truecm
\hangafter = 1
Turyshev S.G., Williams J.G., 2007, Int. J. Mod. Phys. D 16, No. 12A, 2165

\end{document}